\documentclass[twocolumn,showpacs,preprintnumbers,amsmath,amssymb,prb]{revtex4}
\usepackage{graphicx}
\usepackage{dcolumn}
\usepackage{bm}
\usepackage{epstopdf}

\begin{document}

\title{
Equilibrium Dynamics of the Sub-Ohmic Spin-boson Model Under Bias }

\author{Da-Chuan Zheng}
\email{zhengdachuan@ruc.edu.cn}
\affiliation{Department of Physics, Renmin University of China, 100872 Beijing, China}
\author{Ning-Hua Tong}
\email{nhtong@ruc.edu.cn}
\affiliation{Department of Physics, Renmin University of China, 100872 Beijing, China}
\date{\today}

\begin{abstract}
Using the bosonic numerical renormalization group method, we studied the equilibrium dynamical correlation function $C(\omega)$ of the spin operator $\sigma_z$ for the biased sub-Ohmic spin-boson model. The small-$\omega$ behavior $C(\omega) \propto \omega^s$ is found to be universal and independent of the bias $\epsilon$ and the coupling strength $\alpha$ (except at the quantum critical point $\alpha =\alpha_c$ and $\epsilon=0$). Our NRG data also show $C(\omega) \propto \chi^{2}\omega^{s}$ for a wide range of parameters, including the biased strong coupling regime ($\epsilon \neq 0$ and $\alpha > \alpha_c$), supporting the general validity of the Shiba relation. Close to the quantum critical point $\alpha_c$, the dependence of $C(\omega)$ on $\alpha$ and $\epsilon$ is understood in terms of the competition between $\epsilon$ and the crossover energy scale $\omega_{0}^{\ast}$ of the unbiased case. $C(\omega)$ is stable with respect to $\epsilon$ for $\epsilon \ll \epsilon^{\ast}$. For $\epsilon \gg \epsilon^{\ast}$, it is suppressed by $\epsilon$ in the low frequency regime. We establish that $\epsilon^{\ast} \propto (\omega_0^{\ast})^{1/\theta}$ holds for all sub-Ohmic regime $0 \leqslant s < 1$, with $\theta=2/(3s)$ for $0 < s \leqslant 1/2$ and $\theta = 2/(1+s)$ for $1/2 < s < 1$. The variation of $C(\omega)$ with $\alpha$ and $\epsilon$ is summarized into a crossover phase diagram on the $\alpha-\epsilon$ plane.

\end{abstract}
\pacs{05.10.Cc, 05.30.Jp, 64.70.Tg, 75.20.Hr}

\keywords{ spin-boson mode, numerical renormalization group, quantum phase transition, dynamical correlation function}

\maketitle

\begin{section}{Introduction }

The spin-boson model (SBM) is one of the simplest models to describe a quantum two-level system coupled to the environmental noise \cite{Leggett,Weiss1}. It has realizations in various fields of physics, including the superconducting qubit \cite{Shnirman}, mesoscopic metal ring penetrated by an Aharonov-Bohm flux \cite{Tong1}, ultraslow glass dynamics \cite{Rosenberg}, heavy fermion metals \cite{Si}, and nanomechanical oscillators \cite{Seoanez}, {\it etc.}. In this model, the environmental noise is represented by a bath of harmonic oscillators with the low frequency spectral function characterized by $J(\omega) \propto \alpha\omega^s$.  Here $\alpha$ is the coupling constant and $s$ is a dimensionless parameter that discriminates the Ohmic ($s=1$), super-Ohmic ($s>1$), and sub-Ohmic ($0\leqslant s <1$) bath. 

The SBM has been studied in detail for the Ohmic and weak-coupling regime to understand the quantum decoherence and dissipation, which are the key issues for making a long-lived quantum bit \cite{Shnirman}, chemical reaction \cite{Mulhbacher}, as well as in the optical absorption in quantum dot \cite{Borri}. The SBM without a bias field also contains non-trivial quantum phase transitions (QPTs) between a weak-coupling delocalized phase and a strong-coupling localized phase for $0 \leqslant s \leqslant 1$. It is an established result that a Kosterlitz-Thouless type QPT exists for $s=1$ and the ground state is always delocalized for $s>1$ \cite{Leggett,Weiss1}.
In the sub-Ohmic regime $0 \leqslant s <1$ \cite{Leggett,Weiss1,Kehrein}, numerical methods such as numerical renormalization group (NRG) \cite{Bulla1,Guo}, quantum Monte Carlo (QMC) \cite{Winter}, and exact diagonalization \cite{Alvermann} show that a continuous QPT occurs at certain critical coupling $\alpha_c$, at which both the static and the dynamic quantities show critical behavior \cite{Chin, Anders1, Lu, Hur1, Kast, Liu, Yao, Nalbach, Florens1}. 

For the study of both decoherence and QPT, the equilibrium dynamical correlation function is an important quantity. Most of the previous studies concentrate on the case of zero bias and weak coupling, {\it i.e.}, $\epsilon=0$ and $\alpha < \alpha_c$. In this regime, the $\sigma_z - \sigma_z$ correlation function $C(\omega)$ (to be defined below) shows a power law behaviour $C(\omega) \sim \omega^{s}$ in small frequency $\omega \ll \omega_{0}^{\ast}$ and critical behavior $C(\omega) \sim \omega^{-s}$ in the intermediate frequency $\omega_{0}^{\ast} \ll \omega \ll \omega_0$. Here $\omega_0$ is a non-universal high energy scale and $\omega_{0}^{\ast}$ is the crossover energy scale between the delocalized fixed point and the quantum critical fixed point. 

The situation with a finite bias $\epsilon  \neq 0$, however, receives less attention. For $\epsilon \neq 0$, the parity symmetry of the unbiased SBM is broken and hence the localize-delocalize QPT of the sub-Ohmic symmetric SBM no longer exists. Instead, as the coupling strength increases, the ground state changes smoothly from the biased weak-coupling delocalized-like state to the strong-coupling localized-like state with broken symmetry. The strong coupling regime is difficult to described for the perturbation-based theory \cite{Lu}. For the numerical approaches such as NRG \cite{Bulla1} and exact diagonalaization \cite{Alvermann}, a finite bias increases the number of bosons in the ground state and makes an accurate calculation more difficult. As a result, systematic study of the dynamical correlation function for the sub-Ohmic bath under a finite bias, especially in the strong coupling regime $\alpha > \alpha_c$, is still lacking.

In this paper, we use NRG to study the equilibrium state dynamical correlation function of the sub-Ohmic SBM with a finite bias field $\epsilon \neq 0$, with emphasis on the strong-coupling regime $\alpha > \alpha_c$. By carefully extrapolating the boson-state truncation to infinity, we find that except for the exact QPT point, $C(\omega)$ always obeys $\chi^{2} \omega^s$ behavior in the small frequency limit, irrespective of the values of $\epsilon$ and $\alpha$. In the vicinity of the quantum critical point (QCP), $C(\omega)$ is characterized by two different power law regimes: $C(\omega) \propto \omega^{s}$ in the low frequency regime $ \omega \ll \omega^{\ast}$ and $C(\omega) \propto \omega^{-s}$ in the intermediate frequency regime $ \omega^{\ast} \ll \omega \ll \omega_0$, with $\omega_0$ being a non-universal high energy scale. The crossover frequency $\omega^{\ast}$ is tuned by $\epsilon$. For weak bias $\epsilon \ll \epsilon^{\ast}$, $\omega^{\ast} = \omega_{0}^{\ast} \propto |\alpha - \alpha_c|^{z\nu}$ and $C(\omega)$ is not changed significantly from the zero bias case. For strong bias $\epsilon \gg \epsilon^{\ast}$, $\omega^{\ast} \propto \epsilon^{\theta}$ and $C(\omega)$ is suppressed in $\omega \ll \omega^{\ast}$. The behavior of $\omega^{\ast}$ is understood in terms of the competition between the unbiased crossover scale $\omega_0^{\ast}$ and $\epsilon$, and we have $\omega^{\ast} \sim {\text max}\left[ \epsilon^{\theta}, \omega_{0}^{\ast} \right]$. We obtained the critical exponent $\theta$ as a function of $s$. The crossover bias $\epsilon^{\ast}$ is thus determined as $\epsilon^{\ast} = \left( \omega_0^{\ast} \right)^{1/\theta}$. We finally summarize the behavior of $C(\omega)$ in a crossover $\epsilon-\alpha$ phase diagram.

This paper is organized as the following. In section II we introduce SBM and the formalism we used to calculate the equilibrium correlation function with bosonic NRG method. Section III presents results from our NRG study. A conclusion is given in section IV.

\end{section}

\begin{section}{Model and Method }

The Hamiltonian of SBM reads
\begin{eqnarray}
&&  H = -\frac{\Delta}{2} \sigma_x + \frac{\epsilon}{2} \sigma_z + \sum_{i} \omega_{i} a_{i}^{\dagger}a_{i} + \frac{\sigma_z}{2} \ \sum_{i} \lambda_{i}(a_{i}+a_{i}^{\dagger}). \nonumber \\
  &&
\end{eqnarray}
The first two terms describe a two level system with bias $\epsilon$ and tunnelling strength $\Delta$. The bosonic bath is described by the third term, where $\omega_{i}$ is the frequency of the $i$-th boson mode. In the last term, the two level system is coupled to the bosonic bath through $\sigma_z$ and the boson displacement operator. Various experimental realizations of this Hamiltonian have been proposed with tunable parameters $\epsilon$ and $\Delta$ \cite{Tong1,Hur2,Goldstein,Egger}. The influence of bath on the two level system is characterized by the spectral function
\begin{equation}
   J(\omega) = \pi \sum_{i} \lambda_{i}^{2} \delta(\omega -   \omega_{i}).
\end{equation}
In this paper, we use a power law form of $J(\omega)$ with a hard cut-off at $\omega_c$,
\begin{equation}
   J(\omega) = 2\pi\alpha \omega^{s} \omega_{c}^{1-s}, \,\,\,\,\, \,\,\,\, (0 \leqslant \omega  \leqslant   \omega_c).
\end{equation}
We set $\omega_c = 1.0$ as the unit of energy and fix $\Delta=0.1$ to study the dependence of equilibrium dynamics on $\epsilon$ and $\alpha$ for a sub-Ohmic bath.

At $\epsilon=0$, Eq.(1) is invariant under the combined boson and spin parity transformation $U a_{i} U^{-1} = -a_i$ and $U \sigma_{z} U^{-1} = - \sigma_z$. For the sub-Ohmic ($0 \leqslant s<1$) and the Ohmic ($s=1$) baths, a spontaneous breaking of this symmetry may occur in the regime $\alpha > \alpha_c$ and the system enters the localized phase, in which the quantum system is trapped to one of the two states and the local bosons has a finite displacement \cite{Kehrein,Bulla1,Guo,Winter,Alvermann}. This is the delocalize-localize quantum phase transition of SBM. With a finite bias $\epsilon \neq 0$, the above parity symmetry is broken from the outset and the phase transition no longer exists. As $\alpha$ increases, the ground state crosses over smoothly from the weak-coupling to the strong-coupling biased states.

In this paper, we focus on the $\sigma_z-\sigma_z$ dynamical correlation function defined as
\begin{equation}
   C(\omega)=\frac{1}{2\pi} \int_{-\infty}^{+\infty} C(t) \,dt,
\end{equation}
where $C(t)=(1/2)\langle [ \sigma_z(t),\sigma_z(0) ]_{+}\rangle$ and $[\hat{A}, \hat{B} ]_{+}$ is the anti-commutator of $\hat{A}$ and $\hat{B}$. For a non-degenerate ground state, the Lehman representation of the correlation function at $T=0$ is written as
\begin{eqnarray}
   C(\omega)&=& \frac{1}{2} \sum_{n} | \langle 0|\sigma_z|n \rangle |^2 \delta(\omega+ E_{0}-E_{n})   \nonumber \\
   && +  \frac{1}{2} \sum_{n} | \langle n|\sigma_z|0 \rangle |^2 \delta(\omega+ E_{n}- E_{0}) .
\end{eqnarray}
Here $|n \rangle$ and $E_n$ are the $n$-th eigen state and energy of the Hamiltonian, respectively. $C(\omega)$ has the general form $C(\omega)= A \delta(\omega) + C^{\prime}(\omega)$, where $A= |\langle 0| \sigma_z |0 \rangle|^{2}$ and $|0 \rangle$ is the ground state. 
It is an even function of $\omega$ and fulfils the sum rule
\begin{equation}
   \int_{-\infty}^{+\infty} C(\omega)  \, d \omega = 1.
\end{equation}

The bosonic NRG method is regarded as one of the most accurate numerical techniques for studying SBM due to its non-perturbative nature and the applicability in the whole range of parameters \cite{Wilson,Bulla1,Bulla2}. The success of NRG relies on the energy scale separation due to the logarithmic discretization, and on the RG transformation which is carried out by iterative diagonalization in each energy shell. It is technically composed of three steps: logarithmic discretization, transforming the Hamiltonian into a semi-infinite chain, and the iterative diagonalization. In general, the errors in the NRG calculation come from two sources. One is the approximation of using one bath mode to represent each energy shell, whic is controlled by the logarithmic discretization parameter $\Lambda \geqslant 1$. The other is the truncation of the energy spectrum after each diagonalization to overcome the exponential increase of the Hilbert space, which is controlled by the number of kept states $M_s$. For the bosonic NRG, an additional source of error is the truncation of infinite dimensional Hilbert space of each boson mode into $N_b$ states on the occupation basis. Exact results are obtained in the limit $\Lambda=1$, $M_s= \infty$, and $N_b = \infty$. 

In practice, however, one cannot do calculation directly at the above limit and extrapolating NRG data to the above limit could be difficult. In the strong coupling regime with a bias, the ground state has a large boson number and hence requires a large $N_b$ for an accurate calculation. A large $N_b$, however, will increase the truncation error because in NRG calculation only the lowest $1/N_b$ fraction of the eigen states of previous energy shell are kept for constructing the Hamiltonian of the lower energy shell. To keep the accuracy, one has to use a larger $\Lambda$ to increase the energy separation of successive energy shells. This in turn will lead to a larger discretization error of the bath. Fortunately, we found that a larger $\Lambda$ only induces quantitative changes to the NRG results and the qualitative conclusion can still be obtained reliably. Therefore, using a large $\Lambda = 2  \sim 10$, a moderate $M_s \sim 100 \sim 200$, and $N_b = 12 \sim 60$, we can obtain qualitatively correct results for the biased SBM in the strong coupling regime. Combining the knowledge gained from the finite size scaling method of $N_b$ (Ref.~\onlinecite{Hou1,Tong2}), we checked that our final conclusions are stable when extrapolated to the exact limit. To calculate $C(\omega)$, we use the patching method of Bulla~\cite{Bulla3} with which the sum rule Eq.(6) is fulfilled reasonably well. The discrete $\delta$ peaks obtained are broadened with a Gaussian function on the logarithmic scale \cite{Bulla3},
\begin{equation}
   \delta(\omega - \omega_n) \rightarrow \frac{e^{-b^2/4}}{b\omega_n\sqrt{\pi}} \exp { \left[ -\frac{( \ln{\omega}-\ln{\omega_n} )^2}{b^2} \right] }.
\end{equation}
We choose the broadening parameter $b=0.7$ for $\Lambda < 4$  and $b=2.0$ for $\Lambda \geqslant 4$. 

\end{section}

\begin{section}{Results }

In this work, we study the dynamical correlation function $C(\omega)$ for all range of $\alpha$ and $\epsilon \neq 0$. As $\epsilon$ tends to zero, the strong coupling phase of $\alpha > \alpha_c$ is continuously connected to the localized state at $\epsilon =0$ with spontaneously broken symmetry. At $\epsilon=0$ and $\alpha > \alpha_c$, the localized state without symmetry breaking is special in that the ground state has a two-fold degeneracy and it is unstable in the thermodynamical limit \cite{Cao}. In this paper we hence confine our study to the symmetry-broken phase (either spontaneously or by a finite bias) and do not study that special case.

\begin{subsection}{$\omega^s$-behavior in the small frequency regime}

\begin{figure}
\begin{center}
  \includegraphics[width=280pt, height=200pt, angle=0]{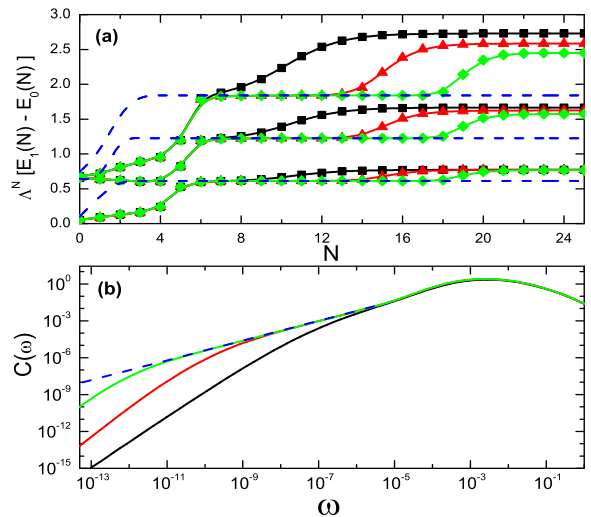}
  \vspace*{-1.0cm}
\end{center}
  \caption{(color online) NRG flow diagrams and $C(\omega)$ at $s=0.8$, $\Delta=0.1$, $\alpha=0.5 > \alpha_c$, and $\epsilon=10^{-3}$. (a) Flow diagrams obtained using $N_b=8$, $18$, and $40$ (symbols with eye-guiding lines, from left to right). The dashed lines are the flow of $\alpha=0$ and $\epsilon=0$. (b) Dynamical correlation function $C(\omega)$ for $N_{b}=8$, $18$, and $40$ (solid lines, from bottom to top), and $N_{b}=\infty $ from extrapolation (dashed lines). Fitting the small $\omega$ regime, we obtain the exponents $y_{l}=2.07$ for $N_b=8$, and $y_{h} = 0.799$ for $N_b = \infty$. The $\delta(\omega)$ peak is not shown. NRG parameters are $\Lambda=4.0$ and $M_s=120$.
}\label{fig1}
\end{figure}
\begin{figure}
\vspace*{-1.0cm}
\begin{center}
  \includegraphics[width=320pt, height=240pt, angle=0]{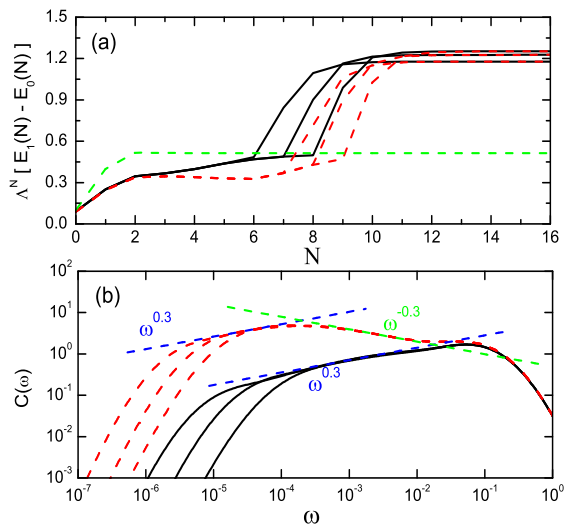}
  \vspace*{-1.0cm}
\end{center}
  \caption{(color online) NRG flow diagrams and $C(\omega)$ at $s=0.3$, $\Delta=0.1$, $\alpha=0.045 > \alpha_c$. $\epsilon=10^{-2}$ (solid lines) and $\epsilon=10^{-3}$ (dashed lines). (a) Flow diagrams obtained using $N_b=12$, $27$, and $60$ (from left to right). The horizontal dashed line are the flow of $\alpha=0$ and $\epsilon=0$. (b) $C(\omega)$ for $N_b=12$, $27$, and $60$ (from bottom to top). The functions $\omega^{0.3}$ and $\omega^{-0.3}$ are marked by the dashed straight lines for guiding the eye. The $\delta(\omega)$ peak is not shown. NRG parameters are $\Lambda=4.0$ and $M_s=120$.
}\label{fig2}
\end{figure}

For the Ohmic ($s=1$) and the super-Ohmic ($s>1$) baths, the small frequency behavior of $C(\omega)$ in the delocalized phase obeys the Shiba relation \cite{Weiss2}, which in our notation reads\cite{Bulla1}
\begin{equation}
    C(w)=  \frac{\alpha}{4} {\chi}^2 \omega^{s},  \,\,\,\,\,\,\,\,\,\,\,\,\,\,\, (\omega \to 0).
\end{equation}
Here $\chi \equiv 2 \partial \langle \sigma_z \rangle / \partial \epsilon|_{\epsilon=0}$ is the local spin susceptibility. The proof in Ref.~\onlinecite{Weiss2} applies also to the case with a finite bias, but not to the localized phase at $\alpha > \alpha_c$ for the Ohmic bath. This exact relation was used to test the quality of various approximate results for SBM \cite{Florens1,Egger2,Costi,Volker,Keil,Bulla1}.

There have been attempts to generalize the Shiba relation to (i) sub-Ohmic bath, and (ii) with finite bias, and (iii) strong coupling regime $\alpha > \alpha_c$. Such a generalized relation, if exists, would imply an universal long time behavior $C(t) \propto t^{-(1+s)}$ for any $s$ values, coupling strength $\alpha$ (except at the critical point $\alpha = \alpha_c$ and $\epsilon=0$), and spin parameters $\Delta$ and $\epsilon$. Up to now these activities received partial success only. Florens and others \cite{Florens2} proved Eq.(8) for the sub-Ohmic bath, based on an exact relation from the perturbation theory in Majorana representation. However, their proof applies only to delocalized phase at the symmetry point $\epsilon=0$. The numerical results from the approximate perturbation theory built on an unitary transformation fulfils Eq.(8) exactly for the sub-Ohmic bath with \cite{Gan} or without \cite{Lu} bias in the weak to intermediate coupling. In the strong coupling regime $\alpha > \alpha_c$ the fulfilment is good but not exact. This leaves the question open whether Shiba relation holds in the biased strong coupling regime of the sub-Ohmic SBM.

As stated above, the inherent truncation errors of NRG hinder it from the quantitative confirmation/falsification of the generalized Shiba relation. However, qualitatively, it is possible to check the factors on the right-hand side of Eq.(8) each for a time. In this section, we will show that $C(\omega) \propto \omega^{s}$ holds for the sub-Ohmic SBM with general $\alpha$ and $\epsilon$. In Fig.6, data are presented to further support that $C(\omega)/\omega^{s} \propto \chi^{2}$ for general $\alpha$ and $\epsilon$ values.

In Fig.1, we first present the $N_b$-dependence of $C(\omega)$ obtained from NRG under bias, in order to gauge the choice of $N_b$ in our study. The energy flow (Fig1.(a)) and  $C(\omega)$ (Fig1.(b)) are shown for a sub-Ohmic bath $s=0.8$ in the strong coupling regime $\alpha=0.5 > \alpha_c$, with a finite bias $\epsilon=10^{-3}$. We use a series of $N_b$ values from $N_b=8$ to $N_b=40$ with $\Lambda = 4$ and $M_s=120$. In Fig.1(a), for a given $N_b$, the excitation energies $\Lambda^{N}\left[ E_{i}(N) - E_{0}(N) \right]$ ($i=1, 2, 3$) first flow to an unstable fixed point (with $\Lambda^{N}\left[ E_{1}(N) - E_{0}(N) \right] \approx 0.61$) in the small $N$ regime. After a crossover $N^{\ast}$, they flow to the stable fixed point ( with $\Lambda^{N}\left[ E_{1}(N) - E_{0}(N) \right] \approx 0.77$) in the large $N$ limit. 
We found that the unstable fixed point in the small $N$ regime is same as the weak-coupling fixed point of the SBM in the non-bias case (dashed lines obtained using $\alpha=0$ and $\epsilon=0$). With $N_b$ increasing geometrically, the crossover $N^{\ast}$ increases linearly, showing that the associated energy scale $T^{\ast} \sim \Lambda^{-N^{\ast}}$ decreases to zero as a negative power of $N_b$. This supports that the unstable fixed point (dashed lines in Fig.1(a)) will extend to infinitely large $N$ in the limit $N_b = \infty$ and it is the true fixed point of the biased SBM. In this biased fixed point, $\epsilon$ flows to infinity and the spin is effectively decoupled from the bath, leading to the same excitation levels as the free boson chain.

Fig.1(b) shows the corresponding $C(\omega)$, which has two different power-law regimes: $C(\omega) \sim \omega^{y_l}$ in the low frequency regime $\omega \ll \omega^{\ast}$, and $C(\omega) \sim \omega^{y_h}$ in the high frequency regime $\omega \gg \omega^{\ast}$. NRG data for $s=0.8$ gives $y_l=2.07$ and $y_{h}=0.799$. With increasing $N_b$, the crossover frequency $\omega^{\ast}$ decreases to zero as a negative power of $N_b$, proving that $C(\omega) \sim \omega^{y_h}$ is the correct result of the biased SBM (dashed line in Fig.1(b)). Our results for other $s$ values in $1/2 \leqslant s \leq 1$ agree with $y_h = s$ within an error of $2 \%$ (not shown). 

For $s<1/2$, the NRG results converge much more difficult with increasing $N_b$. In Fig.2, we show the strong coupling data ($\alpha=0.045 > \alpha_c$) for $s=0.3$. Both the energy flow and $C(\omega)$ have $N_b$-dependent crossover scales. With $N_b$ increasing to $60$, a clear trend can be seen that the $N_b$-converged levels flow towards the free boson energy levels (Fig.2(a)). For both $\epsilon=10^{-2}$ and $\epsilon=10^{-3}$, the section of $C(\omega)$ with $\omega^{s}$ behavior extends to smaller $\omega$ with increasing $N_b$ (Fig.2(b)). The shift of $C(\omega)$ with $N_b$ has an apparent scaling form and an analysis for $N_b$ could be carried out to extract the correct exponent, as done in Ref.~\onlinecite{Tong2}. Here in Fig.2(b), we only mark out the expected asymptotic $\omega^{s}$ line for guiding the eye. 
With this understanding of the $N_b$-dependence of $C(\omega)$, in the rest part of this paper, we only show the physically correct results $C(\omega > \omega^{\ast})$ obtained using sufficiently large $N_b$.

\begin{figure}
\begin{center}
  \includegraphics[width=300pt, height=210pt, angle=0]{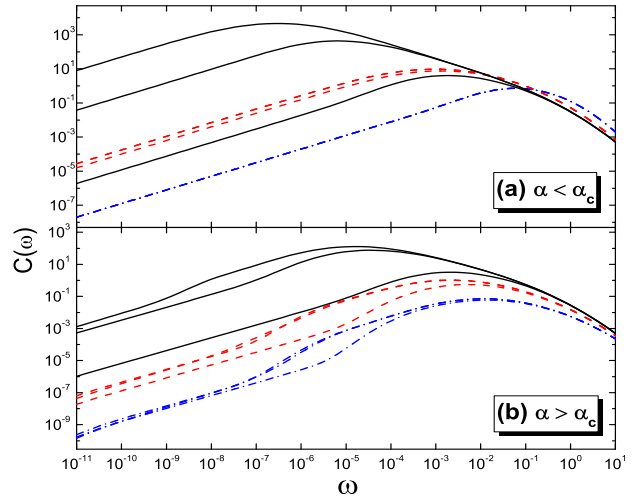}
  \vspace*{-1.0cm}
\end{center}
  \caption{(color online) $C(\omega)$ for various $\alpha$ and $\epsilon$ at $s=0.8$, $\Delta=0.1$. (a) $\alpha < \alpha_c \approx 0.482$: $\alpha=0.48$ (solid lines), $0.4$ (dashed lines), and $0.1$ (dash-dotted lines). (b) $\alpha > \alpha_c$: $\alpha=0.49$ (solid lines), $0.57$ (dashed lines), and $0.7$ (dash-dotted lines). For each $\alpha$ value, from top to bottom are curves for $\epsilon=0$, $10^{-5}$, and $10^{-3}$. All curves show $C(\omega) \propto \omega^s$ behavior in the small $\omega$ limit. NRG parameters are $\Lambda=4$, $M_s=120$, and $N_b=60$.
}  \label{fig3}
\end{figure}

In Fig.3, we investigate the influence of $\epsilon$ on $C(\omega)$ for $\alpha$ values ranging from weak coupling $\alpha < \alpha_c$ (Fig.3(a)) to strong coupling $\alpha > \alpha_c$ (Fig.3(b)) for $s=0.8$ and $\Delta=0.1$. It is seen that $C(\omega) \sim \omega^{s}$ in the small frequency limit for all parameters. 
For the small coupling $\alpha=0.1$, $C(\omega)$ does not change for $\epsilon$ as large as $10^{-3}$. While for a moderate coupling $\alpha=0.4$, a slight downward shift is observed between $\epsilon=10^{-3}$ and $\epsilon=10^{-5}$. $C(\omega)$ is most sensitive to $\epsilon$ when $\alpha$ is close to $\alpha_c \approx 0.482$.

The universal $C(\omega) \propto \omega^{s}$ behavior observed so far in Figs.1, 2, and 3 has a natural understanding. When the parity symmetry already broken by a finite bias, the ground state of SBM can be tuned continuously on the $\alpha-\epsilon$ plane, going from a delocalized state at $\alpha < \alpha_c$, $\epsilon=0^{+}$, through a half circle in finite $\epsilon$ region, to the symmetry-spontaneously-broken state at $\alpha > \alpha_c$, $\epsilon=0^{+}$, without passing the critical point. Therefore, the ground state has same nature and no qualitative change is expected in the small $\omega$ limit of $C(\omega)$. Indeed, our NRG results for different $s$, $\alpha$, and $\epsilon$ confirm that $C(\omega) \propto \omega^{s}$ is a universal feature of SBM. 

\begin{figure}
\begin{center}
  \includegraphics[width=320pt, height=230pt, angle=0]{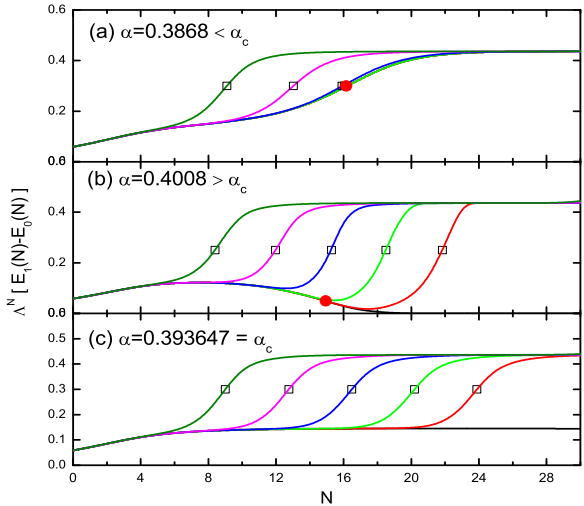}
  \vspace*{-2.0cm}
\end{center}
  \caption{(color online) (a)-(c): the flow diagrams for three different $\alpha$ values. For each $\alpha$, from right to left $\epsilon=0$, $10^{-7}$, $10^{-6}$, $10^{-5}$, $10^{-4}$, and $10^{-3}$. The empty squares make out the crossover $N^{\ast}$ for the case $\epsilon >0$. The solid circle marks out the crossover scale in the unbiased case $N_{0}^{\ast}$. In (c), $N_{0}^{\ast} = \infty$. NRG parameters are $\Lambda=2.0$, $M_s=200$, and $N_b=16$.
}     \label{Fig4}
\end{figure}
\begin{figure}
\begin{center}
  \includegraphics[width=320pt, height=230pt, angle=0]{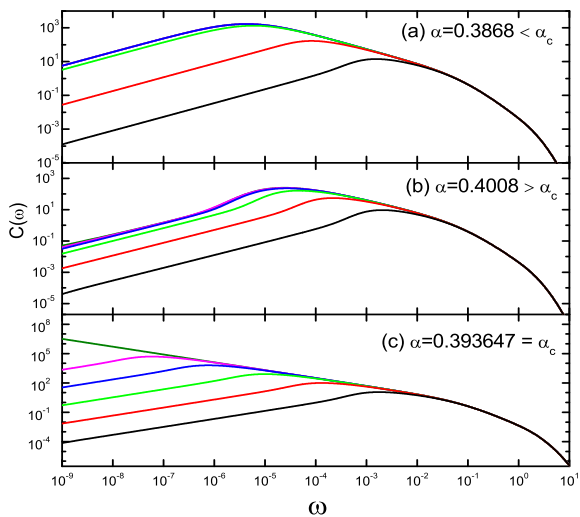}
  \vspace*{-1.2cm}
\end{center}
  \caption{(color online)  (a)-(c): $C(\omega)$ for the same $\alpha$ values. For each $\alpha$,  from top to bottom, $\epsilon=0$, $10^{-7}$, $10^{-6}$, $10^{-5}$, $10^{-4}$, and  $10^{-3}$. NRG parameters are same as Fig.4.
}    \label{Fig5}
\end{figure}

In the strong coupling case shown in Fig.3(b), the same tendency is observed, {i.e.}, with increasing $\epsilon$, $C(\omega)$ shifts downwards and the most prominant change occurs near $\alpha_c$. The suppression of $C(\omega)$ can be understood as the weight transfer: with increasing $\epsilon$, $|\langle \sigma_z \rangle|$ increases and so does the weight of the zero-frequency $\delta$ peak. Due to the sum rule,  $C(\omega > 0)$ decreases uniformly. Comparing Fig.3(a) and (b), one sees that a broad peak forms for $\alpha > \alpha_c$ and $\epsilon \sim 10^{-3}$.

\end{subsection}

\begin{subsection}{Equilibrium Dynamics near QCP}

In the above section, we established the universal $\omega^{s}$-behaviour of $C(\omega)$ in the full parameter space of SBM. In this section, we focus on the parameter regime near the QCP $\alpha \sim \alpha_c$ and study the critical properties of $C(\omega)$ under a bias.
For the unbiased SBM, $T_{0}^{\ast} \propto |\alpha - \alpha_c|^{z\nu}$ is the only energy scale that controls the crossover between the stable fixed points (localized and delocalized fixed points) and the critical fixed point. Here $\nu$ is the critical exponent of correlation length and $z=1$ is the dynamical critical exponent. $T_{0}^{\ast}$ plays an important role in the temperature dependence of physical quantities close to QCP. At zero temperature, it also appears in the dynamical correlation function in the delocalized phase:  $C(\omega) \sim \omega^{s}$ for $\omega \ll \omega_{0}^{\ast}$ and $C(\omega) \sim \omega^{-s}$ for $\omega \gg \omega_{0}^{\ast}$, and $\omega_0^{\ast} = T_{0}^{\ast}$.

For the biased case, $\epsilon >0$ is another energy scale that influences the crossover between different behaviors of $C(\omega)$. Our NRG results for $C(\omega)$ at different $\alpha$ and $\epsilon$ can be understood in terms of the competition between $\epsilon$ and $T_{0}^{\ast}$. 
Due to the difficulty of $N_b$ convergence for $0 \leqslant s < 1/2$, here we show NRG data for $s=0.8$, representing a typical case for $1/2 < s < 1$. Our conclusion also applies to $0 \leqslant s < 1/2$, as will be discussed below.

In Fig.4 and Fig.5, we show the flow diagrams (Fig.4) and $C(\omega)$ (Fig.5) for $\alpha=0.3868 < \alpha_c$ (a), $\alpha=0.4008 > \alpha_c$ (b), and $\alpha=0.393647 \approx \alpha_c$ (c), respectively. For each $\alpha$, $\epsilon$ varies from zero to $10^{-3}$. The purpose is to observe the influence of $\epsilon$ in the delocalized, localzied, and critical phases. Fig.4(a) and Fig.5(a) are for $\alpha < \alpha_c$. At $\epsilon=0$, the energy flow in Fig.4(a) has a crossover at around $N_{0}^{\ast} \approx 16$ (solid circle), from the critical fixed point to the weak-coupling one. As $\epsilon$ increases from zero, the flow does not change for $\epsilon < 10^{-5}$ and the crossover $N^{\ast}$ (empty squares) begins to decreases only for $\epsilon > 10^{-4}$. The corresponding evolution of $C(\omega)$ is shown in Fig.5(a). For $\epsilon < 10^{-5}$, $C(\omega) \propto \omega^{-s}$ for $\omega \gg \omega_{0}^{\ast}= T_{0}^{\ast}$ and $C(\omega) \propto \omega^{s}$ for $\omega \ll \omega_{0}^{\ast}$. The crossover scale of the symmetric SBM is given by $\omega_{0}^{\ast} =  \Lambda^{-N_{0}^{\ast}} \approx 10^{-5}$. With increasing $\epsilon$, the crossover frequency $\omega^{\ast}$ increases and  $\omega^{\ast} > \omega_{0}^{\ast}$ occurs only when $\epsilon > 10^{-4}$, corresponding to the occurrence of $N^{\ast} < N_0^{\ast}$ in the flow diagram. A crossover scale $\epsilon^{\ast}$ can be defined as such that for $\epsilon > \epsilon^{\ast}$, $\omega^{\ast}$ becomes significantly larger than $\omega_{0}^{\ast}$, or equivalently, $N^{\ast} < N_0^{\ast}$. For $\epsilon \ll \epsilon^{\ast}$, $C(\omega)$ stays same as the symmetric case ($\epsilon=0$) and $\omega^{\ast} = T_{0}^{\ast}$. For $\epsilon \gg \epsilon^{\ast}$, $C(\omega)$ is suppressed in the low frequency regime and $\omega^{\ast} > T_{0}^{\ast}$ is set by $\epsilon$.

Fig.4(b) and Fig.5(b) show the influence of $\epsilon$ in the localized phase. In Fig.4(b), the excitation energy level at $\epsilon=0$ flows from the critical fixed point towards a two-fold degenerate fixed point, with the critical-to-localize crossover around $N^{\ast} = N_{0}^{\ast}=15$ (solid circle). Now we study the change of $N^{\ast}$ under bias (empty squares). With a vanishingly small $\epsilon$, the degeneracy is lifted and the excited energy level has an upturn at an arbitrarily large $N^{\ast} \gg N_{0}^{\ast}$, showing a crossover from the localized symmetric state to the biased state. We find that the crossover energy scale $T^{\ast} \equiv \Lambda^{-N^{\ast}} \propto \epsilon$. With further increasing $\epsilon$, the upturn moves to the left and for $\epsilon > \epsilon^{\ast}$, $N^{\ast} < N_{0}^{\ast}$ occurs, which means that $N^{\ast}$ is now the crossover from the critical fixed point to the biased fixed point. NRG data give the critical-to-biased crossover $T^{\ast} \propto \epsilon^{\theta}$. The critical exponent $\theta$ will be discussed with Fig.5(c).

In Fig.5(b), the evolution of $C(\omega)$ with $\epsilon$ is shown, which looks similar to the case $\alpha < \alpha_c$. That is, $C(\omega)$ does not change much with $\epsilon$ for $\epsilon \ll \epsilon^{\ast}$ and begins to be suppressed for $\epsilon \gg \epsilon^{\ast}$. It is noted that for a given $\epsilon < \epsilon^{\ast}$, the degenerate-to-biased crossover at $N^{\ast} > N_{0}^{\ast}$ in the energy flow  has no correspondence in $C(\omega)$: the latter has a perfect $\omega^{s}$ behavior at $\omega \sim \omega^{\ast} = \Lambda^{-N^{\ast}}$. The crossover in the energy flow will only show up in the temperature dependence of $C(\omega)$.

\begin{figure}
\vspace*{-3.0cm}
\begin{center}
\includegraphics[width=400pt, height=300pt, angle=0]{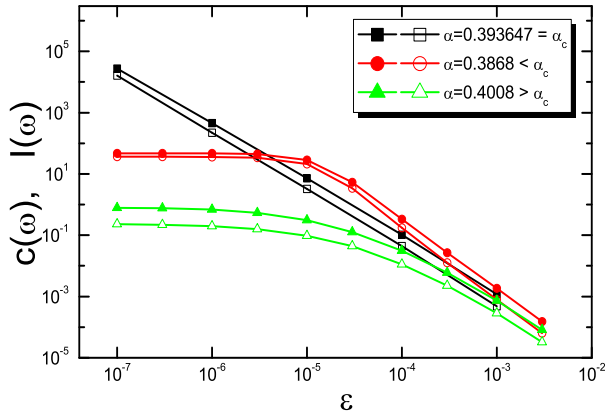}
\vspace*{-2.5cm}
\end{center}
  \caption{(color online)   Checking the Shiba relation at $s=0.8$ and $\Delta=0.1$. $C(\omega)$ (empty symbols) is compared to $I(\omega) \equiv \alpha \chi^{2} \omega^{s} / 4$ (solid symbols) for various $\alpha$ and $\epsilon$ values. Here $\omega = 1.01561 \times 10^{-8}$. NRG parameters are $\Lambda=2.0$, $M_{s}=200$, and $N_b=16$.
}     \label{Fig6}
\end{figure}

Fig.4(c) and Fig.5(c) show the the influence of $\epsilon$ on the flow and $C(\omega)$ in the critical regime $\alpha \approx \alpha_c$. Since $T_{0}^{\ast}=0$ at this point, $\epsilon$ is the only energy scale that controls the critical-to-biased crossover in the energy flow and $C(\omega)$. Especially, Fig.5(c) shows that $C(\omega) \propto \omega^{-s}$ for $\omega \gg \omega^{\ast}$ and $C(\omega) \propto \omega^{s}$ for $\omega \ll \omega^{\ast}$. We define an critical exponent $\theta$ as 
\begin{equation}
   \omega^{\ast}(\alpha = \alpha_c, \epsilon) \propto \epsilon^{\theta}.
\end{equation}
For $s=0.8$, the fitted exponent from Fig.5(c) is $\theta = 1.12$. Note that the same dependence of $\omega^{\ast}$ on $\epsilon$ applies to the $\alpha > \alpha_c$ and $\alpha < \alpha_c$ cases in the regime $\epsilon \gg \epsilon^{\ast}$, which are shown in (a) and (b) of Figs.4 and 5.

The NRG result in Fig.5(c) shows that for $\alpha = \alpha_c$, $C(\omega \gg \omega^{\ast}) \propto c \omega^{-s}$ with an $\epsilon$-independent factor $c$. Combining this observation with the assumption that the Shiba relation Eq.(8) holds at $\alpha = \alpha_c$ and $\epsilon >0$ (which will be discussed in Fig.6 below), we can derive $\theta$ by equating the small and the large frequency expression at $\omega = \omega^{\ast}$, giving
\begin{equation}
   \left( \omega^{\ast} \right)^{-s} \sim \alpha_c [\chi(\alpha_c, \epsilon)]^{2} \left( \omega^{\ast} \right)^{s}.
\end{equation}
Employing the critical behavior $\chi(\alpha_c) \propto \epsilon^{1/\delta -1}$, one obtains $\omega^{\ast} \propto \epsilon^{(\delta-1)/(\delta s)}$, giving $\theta = (\delta-1)/(\delta s)$. From the exact expression $\delta=3$ for $0 < s < 1/2$ and $\delta = (1+s)/(1-s)$ for $1/2 \leqslant s < 1$, we obtain
\begin{eqnarray}
  \theta = \left\{
\begin{array}{lll}
  \frac{2}{3s},   \,\, \,\,\, \,\,  & (0 < s \leqslant \frac{1}{2} ); \\
& \\
  \frac{2}{1+s},   \,\,\,\,\, \,\,  & ( \frac{1}{2} < s <1).
\end{array} \right. 
\end{eqnarray}
For $s=0.8$, this expression gives $\theta=1.111$, which agree well with the NRG result $1.12$. 

Having shown that $C(\omega) \propto \omega^{s}$ ($\omega \to 0$) for general $\alpha$ and $\epsilon$ values in the previous section, we now check the Shiba relation Eq.(8) at $s=0.8$ using a fixed small frequency $\omega = 1.01561 \times 10^{-8}$. Fig.6 shows $C(\omega)$ and the right-hand side of Eq.(8) $I(\omega) \equiv \alpha \chi^{2} \omega^{s}/4$ as functions of $\epsilon$, for the same $\alpha$ values as in Figs.4 and 5. We find qualitative agreement between them for a wide range of $\epsilon$.
For $\alpha > \alpha_c$ ($\alpha < \alpha_c$), $C(\omega)$ shows the crossover from critical (power law dependence on $\epsilon$) for $\epsilon \gg \epsilon^{\ast}$ to the localized-like (delocalized-like) behavior (being constant) for $\epsilon \ll \epsilon^{\ast}$, as expected. For $\alpha=\alpha_c$, a power-law dependence is observed, with the fitted exponent $-1.88$ (for $C(\omega)$) and $-1.84$ (for $I(\omega)$), respectively, in reasonable agreement with the exact $2(1/\delta -1) = -1.78$. It is notable that although $C(\omega)$ and $I(\omega)$ have more than $5$ decades of variations in the range $\epsilon \in \left[10^{-7} , 10^{-3}\right]$, their ratio $I(\omega)/C(\omega) = 2 \sim 3$ does not change much. This result is a support to the relation $C(\omega) \propto \chi^{2} \omega^{s}$. The ratio also depends weakly on $\omega$ due to the slight inaccuracy in the exponent of NRG-produced $C(\omega)$. The uniform deviation from Eq.(8) observed here is more likely due to the error of NRG data than due to the invalidity of Shiba relation. After all, a factor of $3$ is a reasonable level of error in the NRG calculation of dynamical quantities \cite{Bulla1}, considering the logarithmic error, truncation errors, as well as the approximation used to calculate $C(\omega)$ and $\chi$ (Ref. \onlinecite{Bulla3}). It is expected that the agreement can be improved if we increase $M_{s}$ and $N_{b}$, and extrapolate $\Lambda$ to unity, which will not be pursued here. The good agreement of the NRG value of the exponent $\theta$ and Eq.(11) is also consistent with $C(\omega) \propto \chi^{s}\omega^{s}$, since Eq.(11) is derived from the this assumption.

\begin{figure}
\vspace*{-1.0cm}
\begin{center}
\includegraphics[width=320pt, height=250pt, angle=0]{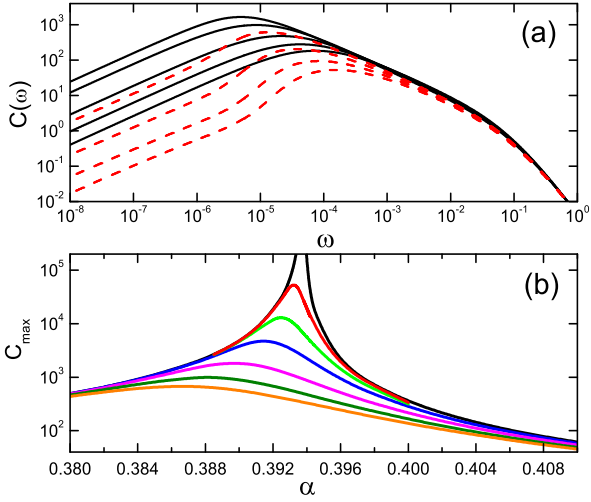}
\vspace*{-1.0cm}
\end{center}
\caption{ (color online)   (a) $C(\omega)$ at $\epsilon=10^{-5}$ and for a series of $\alpha$'s. From top to bottom, $\alpha=0.39$, $0.385$, $0.38$, $0.375$, and $0.37$ (solid lines), and $\alpha=0.395$, $0.40$, $0.405$, and $0.41$ (dashed lines). (b) The maximum value of $C(\omega)$ as functions of $\alpha$ for a series of $\epsilon$. From top to bottom, $\epsilon=0.0$, $2.0 \times 10^{-7}$,  $1.0 \times 10^{-6}$,  $3.0 \times 10^{-5}$,  $9.0 \times 10^{-6}$,  $1.8 \times 10^{-5}$, and $2.7 \times 10^{-5}$. $s=0.8$, $\Delta=0.1$. NRG parameters are $\Lambda=2.0$, $M_{s}=100$, and $N_b=12$.
}     \label{Fig7}
\end{figure}

In Fig.7(a), we fix $\epsilon = 10^{-5}$ and plot $C(\omega)$ for different $\alpha$ values. For $\alpha \ll \alpha_c$, $C(\omega)$ increases with increasing $\alpha$ and the height of the peak reaches a maximum at certain $\alpha < \alpha_c$. With further increasing $\alpha$, $C(\omega)$ begins to decrease uniformly, with the peak height suppressed. In Fig.7(b), we show the maximum of $C(\omega)$ as functions of $\alpha$ for various $\epsilon$'s. From this figure we can extract the $\alpha$ value at which the height of the peak in $C(\omega)$ reaches the maximum.

\begin{figure}
\vspace*{-0.5cm}
\begin{center}
  \includegraphics[width=250pt, height=180pt, angle=0]{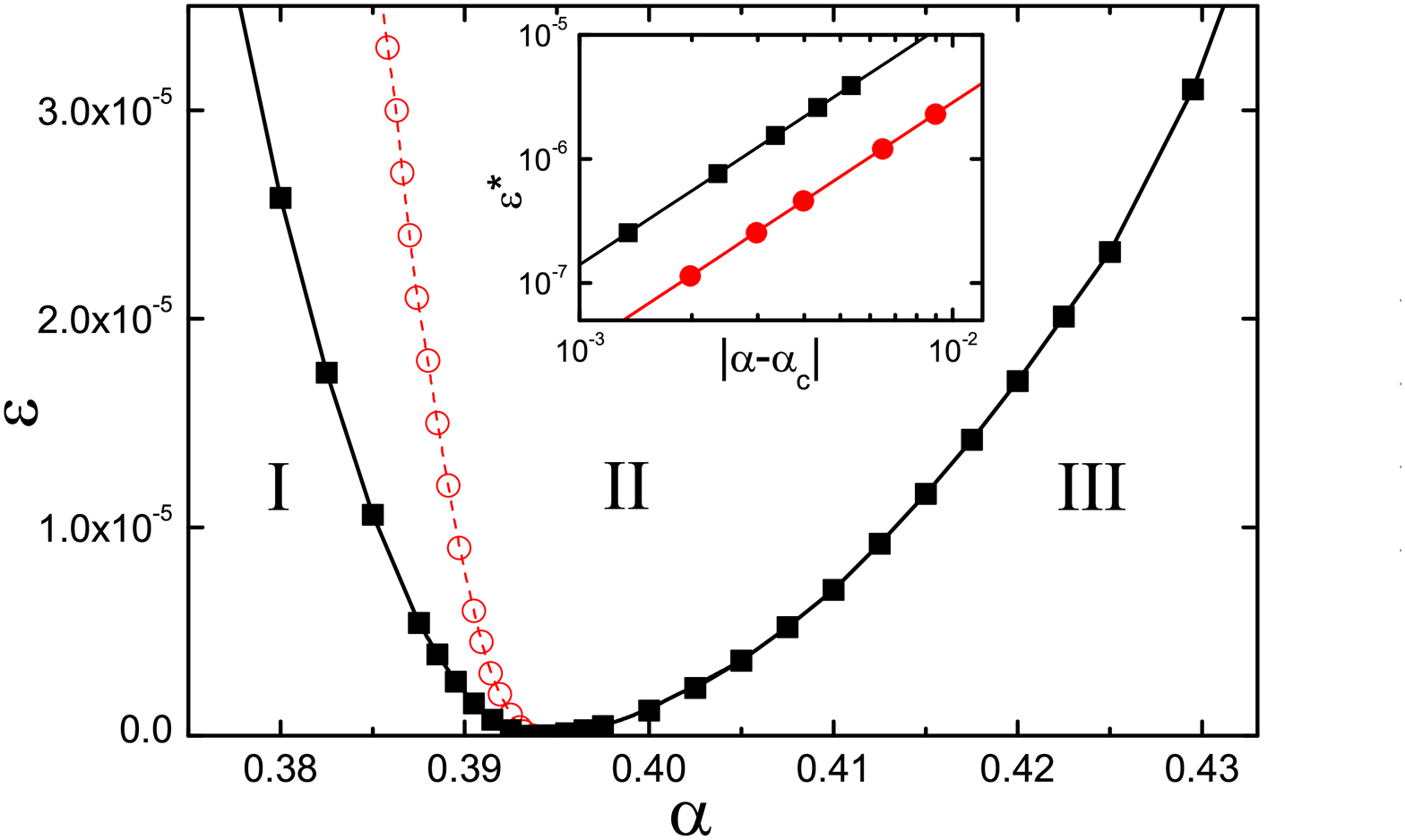}
  \vspace*{-1.0cm}
\end{center}
  \caption{(color online)   Phase diagram of sub-Ohmic SBM near QCP extracted from $C(\omega)$. Region I, II, and III are the delocalized, quantum critical, and the localized phases, respectively. They are separated by a crossover line (solid squares with eye-guiding line) at which the peak height of $C(\omega)$ drops to $90\%$ of its $\epsilon=0$ value. The peak height of $C(\omega)$ reaches maximum at dashed line with empty circles. Inset: log-log plot of $\epsilon^{\ast}$ versus $|\alpha-\alpha_c|$ for $\alpha < \alpha_c$ (solid squares) and $\alpha > \alpha_c$ (solid circles). The fitted exponents are $1.989$ and $1.985$, respectively.
}      \label{Fig8}
\end{figure}

The phase diagram on $\alpha-\epsilon$ plane shown in Fig.8 summarizes our results for $C(\omega)$ near the QCP. Close to $\alpha_c$ and with very small $\epsilon$, three regions I, II, and III of different nature are separated by a crossover line (squares with eye-guiding line). 
Region I and III are continuously connected to the delocalized ($\alpha < \alpha_c$) and localized phase ($\alpha > \alpha_c$) of the symmetric SBM, respectively. They are characterized by $\epsilon^{\ast} \ll T_{0}^{\ast}$. In region I, $\langle \sigma_z \rangle \propto \chi \epsilon$ with $\chi$ being the magnetic susceptibility. In region III, $\langle \sigma_z \rangle \sim -1$ close to the saturate value. For each $\alpha$, the crossover $\epsilon^{\ast}$ in Fig.8 is defined as the $\epsilon$ value at which the maximum of $C(\omega)$ decreases to $90\%$ of its value at $\epsilon=0$.

In region II, $T_{0}^{\ast}$ vanishes near $\alpha_c$ and $\epsilon \gg T_{0}^{\ast}$ becomes the only characteristic energy scale. In this regime, the magnetization $\langle \sigma_z \rangle$ shows the quantum critical behavior as $\langle \sigma_z \rangle \propto \epsilon^{1/\delta}$, with $\delta$ being an critical exponent. 
The empty circles with eye-guiding line marks out the parameter at which $C(\omega)$ has a highest peak. Both the crossover line $\epsilon^{\ast}(\alpha)$ and the peak position can be extracted from the data in Fig.7(b). It is observed that the location of the maximum peak resembles that of the maximum of entanglement entropy \cite{Hur1}. This is not a pure coincident since the crossover peak of $C(\omega)$ reflects a strong fluctuation near the critical point and it is therefore naturally related to the entanglement maximum. 

In the inset of Fig.8, we fit the crossover line close to the QCP in a power law
\begin{equation}
   \epsilon^{\ast} \propto c|\alpha - \alpha_c|^{\eta},
\end{equation}
with the obtained critical exponent $\eta=1.99$ for $s=0.8$. Note that the asymmetry of $\epsilon^{\ast}$ in $\alpha > \alpha_c$ and $\alpha < \alpha_c$ regimes means different pre-factors $c$. Besides using $C(\omega)$ here, the critical behavior of $\epsilon^{\ast}$ can also be determined by the scaling form of $\langle \sigma_z \rangle \left( \alpha, \epsilon, \Delta \right)$ obtained from the $N_b$ scaling analysis of the NRG data \cite{Tong2}. These two approaches give consistent result. For the second approach, the $N_b$ scaling analysis of NRG data gives \cite{Note1}
\begin{equation}
  \langle \sigma_z \rangle(\alpha, \epsilon, \Delta) = m \left[ \left( \frac{\tau}{\Delta^{1-s}} \right)^{\beta}, \left( \frac{\epsilon}{\Delta} \right)^{1/\delta} \right].
\end{equation}
Here $\tau \equiv \alpha - \alpha_c$ and $m(x, y)$ is a two-variable scaling function. By comparing the magnitude of the two variables in Eq.(13), one obtains $\epsilon^{\ast}(\alpha) \propto |\alpha - \alpha_c|^{\beta \delta}$, giving $\eta = \beta \delta$. 
Here, $\beta$ is the critical exponent of the order parameter. It is known that for $0 \leqslant s \leqslant 1/2$, $\beta = 1/2$ and $\delta=3$. For $1/2 < s < 1$, $\beta$ is a function of $s$ whose explicit expression is unknown yet, and $\delta = (1+s)/(1-s)$ (Refs.~\onlinecite{Guo,Winter,Vojta1,Tong2}).
The NRG result $\beta \delta = 1.93$ is in reasonable agreement with the fitted exponent $\eta=1.99$ in the inset of Fig.8, showing the consistency of the static and dynamical approaches.

For $0 \leqslant s < 1/2$, NRG calculation is hindered by the slow convergence with $N_b$ and it is difficult to obtain the quantitative data. However, we note that the scaling form Eq.(13) is obtained from the NRG calculation combined with the $N_b$ scaling analysis for the full sub-Ohmic regime $0 \leqslant s < 1$. It is stronger than the scaling ansatz of free energy which applies only to the regime $1/2 < s < 1$ (Ref.~\onlinecite{Vojta1}). 
The two exponents introduced above, $\theta$ and $\eta$, are not independent. We have $\epsilon^{\ast} \propto |\alpha - \alpha_c|^{ \beta \delta} \sim \left( T^{\ast} \right)^{\beta \delta / (z\nu) }$. This gives $T^{\ast}  \propto \epsilon^{z\nu / (\beta \delta) }$, meaning $\theta = z \nu / \eta$. This is a relation independent of the validity of hyperscaling relation.

\end{subsection}
\end{section}

\begin{section}{ Summary }
In summary, we use the bosonic NRG method to study the equilibrium dynamics of the sub-Ohmic SBM under a bias. We found that the small $\omega$ behvior $C(\omega) \propto \omega^s$ holds for any bias $\epsilon$ and coupling strength $\alpha$ except exactly at the critical point $\alpha = \alpha_c$, $\epsilon=0$, where $C(\omega) \propto \omega^{-s}$. Our results strongly supports that $C(\omega) \propto \chi^{2} \omega^{s}$ in all parameter regimes, including the biased and strong coupling regime. This is in favour of the validity of the generalized Shiba relation in this regime. Close to the QCP, the competition between the energy scale $T^{\ast}_{0}$ and bias $\epsilon$ determines a crossover scale $\epsilon^{\ast}$ which separate the weak-biased regime ($\epsilon \ll \epsilon^{\ast}$) from a strong-biased regime ($\epsilon \gg \epsilon^{\ast}$). Near the QCP, $\epsilon^{\ast} \propto |\alpha-\alpha_c|^{\eta}$ with $\eta = \beta \delta$. These results are summarized in the crossover phase diagram on the $\alpha-\epsilon$ plane.  

\end{section}

\begin{section}{Acknowledgments }

This work is supported by 973 Program of China (2012CB921704), NSFC grant (11374362), Fundamental Research Funds for the Central Universities, and the Research Funds of Renmin University of China 15XNLQ03.

\end{section}

\end{document}